\documentclass[10pt]{iopart}
\usepackage{iopams}   \usepackage{graphics}
\begin{document}

\title[Dynamics of quantum correlations and linear entropy]{Dynamics
of quantum correlations and linear entropy in a multi-qubit-cavity
system}

\author{Alexandra Olaya-Castro\dag \footnote[3]{To whom correspondence
should be addressed (a.olaya@physics.ox.ac.uk)}, Neil F. Johnson\dag,
and Luis Quiroga\ddag }

\address{\dag\ Centre for Quantum Computation and Department of
Physics,  University of Oxford Clarendon Laboratory, Parks Road, OX1
3PU, U.K. }

\address{\ddag\ Departamento de F\'{\i}sica, Universidad de Los Andes,
A.A.4976, Bogot\'a D.C., Colombia }


\begin{abstract} 
We present a theoretical study of the relationship between entanglement and 
entropy in multi-qubit quantum optical systems.  Specifically we
investigate quantitative relations  between the concurrence  and
linear entropy for  a  two-qubit mixed system, implemented as two
two-level atoms  interacting with a single-mode cavity field. The
dynamical evolutions of the  entanglement and entropy, are controlled
via time-dependent cavity-atom  couplings. Our theoretical findings
lead us to propose an alternative  measure of entanglement, which
could be used to develop a much needed correlation  measure for more
general multi-partite quantum systems.
\end{abstract} 



\section{Introduction} 
The distinction between true quantum correlations (entanglement) and 
classical correlations, is a fundamental topic in
physics. Entanglement  plays a central role in quantum  information
(QI) science. However it is a brittle phenomenon: in  any real
experimental situation, the unavoidable  interaction of the quantum
system with its environment results in  decoherence processes, and
will eventually yield mixed quantum states. The  environment of a
single quantum system might be a complex collection of  other quantum
systems with an enormous number of  degrees of freedom, or it might be
as simple as a second single quantum  system. The main properties of
entangled/mixed states have recently been  discussed in Refs.
\cite{ishizaka,kwiat,parkins}. It is thought that  a total increase in
the `mixedness' of a multi-partite system will generally  produce a
decrease in the entanglement.  Nevertheless we believe that a full
understanding of the connection  between entanglement and mixedness,
will require additional insight gained  by considering specific
physical realizations of such multi-partite systems.

Quantum optics, which considers systems of atoms interacting strongly
with  photons (quantum electrodynamics), provides an ideal setting for
the study  of open quantum systems \cite{mabuchi}.  Apart from being
prime candidates  for QI processing \cite{monroe}, such quantum
optical systems are of  fundamental theoretical interest. In
particular, the experimental and  theoretical study of quantum optical
systems may yield insight into the  connections between mixedness and
entanglement. Mixedness, which is  associated with the lack of purity
of a quantum state, is usually  measured by the linear entropy
$M=1-Tr(\rho^2)$ \cite{jaeger} where  $\rho$ is the open quantum
system's density operator. No general measure of  entanglement in
$N$-partite systems is known for $N\geq 3$. For bipartite  systems,
however, entanglement  has been described by several measures.  One of
these is the entanglement of formation, which represents the
asymptotic number of Bell pairs required to  prepare the state using
only local unitary transformations and  classical
communication. Intimately related to this entanglement  measure is the
notion of concurrence. An exact, closed expression for the
concurrence in two-qubit systems, was obtained by Wooters
\cite{wooters}.

In this paper we report on quantitative relations  between the
concurrence  and linear entropy for  a  two-qubit mixed system,
implemented as two two-level atoms  interacting with a single-mode
cavity field. Even when the whole system is  evolving under a single
unitary  evolution, the atoms' surrounding environment (i.e. the
cavity field) can  become entangled with  the atomic subsystem causing
mixedness of the latter. This analysis is  carried out with
generalized  time-dependent cavity-atom couplings, yielding a
generalized dynamical Dicke  model.  Finally we discuss how our
results might extend to multi-particle  systems.
\section{Two-qubit-cavity system} We consider a pair of two-level atoms interacting on resonance
with the quantized mode of an optical field during times shorter than
both the photon and the atomic dipole lifetimes. We systematically
compare the dynamics of  the two-atom entanglement, atom-field
entanglement, and the linear entropy of each of the three constituents
of the system: atom$_1$ ($A_1$), atom$_2$ ($A_2$), and field
($f$). The atoms can interact  with the cavity mode either one at a
time or simultaneously.  In both cases, different time-dependent
couplings between each atom and the field are considered. Comparing
these two situations serves to enhance our understanding  of the
relationships between entanglement and entropy during the {\em
generation}, {\em transfer} or {\em sharing} of entanglement.

In the rotating-wave approximation (RWA) and assuming the atomic
transitions are on  resonance with the frequency of the single-mode
cavity, the Hamiltonian in the  interaction picture is given by
($\hbar=1$)
\begin{eqnarray}
 H_I=f_1(t_1,t,\tau_1)\{a^\dag \sigma_1^{-}+\sigma_1^{+}a\} +
f_2(t_2,t,\tau_2)\{a^\dag \sigma_2^{-}+\sigma_2^{+}a\}
\label{Eq:ham}
\end{eqnarray} where $\sigma_i^{+}=|e_i\rangle \langle g_i|$,
$\sigma_i^{-}=|g_i\rangle \langle e_i|$ with $|e_i\rangle$ and
$|g_i\rangle$ $(i=1,2)$ being the excited and ground states of the
$i$'th atom. Here $a^\dag$ and $a$ are respectively the creation and
annihilation operators for the cavity photons. The time-dependent
coupling of the cavity field with the $i$'th atom, which is injected
at $t_i$ and interacts during a time $\tau_i$, is given by a
time-window function,
\begin{eqnarray} f_i(t_i,t,\tau_i)=[\Theta (t -\tau_i)-\Theta (t -\tau_i-t_i)]\gamma_i(t)
\end{eqnarray} where $\gamma_i(t)$ is the time-dependent atom-field coupling strength. For the
situation in which the second atom flies through the cavity just after
the first one leaves, i.e. $t_2=t_1+\tau_1$, the Hamiltonian
corresponds to  the {\em Dynamical Jaynes-Cummings (DJC)} interaction
for each atom separately.  Entanglement of two and three particles
using this sequential interaction scheme  has been experimentally
achieved \cite{haroche00}. For the case in which $t_1=t_2$  and
$\tau_1=\tau_2$, both atoms interact simultaneously with the cavity
mode,  although they can have different time-dependent couplings. The
situation is described by the generalized two-atom {\em Dynamical
Dicke (DD)} model. Elsewhere we discuss entanglement generation within
this model \cite{ojqpr03}.

For both the DJC model and the DD interaction, the number of
excitations ${\cal N}=a^\dag a + \sum_{i=1,2} \sigma^{+}_i
\sigma^{-}_i$ is a conserved quantity within the RWA. This implies
separable dynamics within subspaces having a prescribed eigenvalue $N$
of ${\cal N}$. For $N=1$ a basis is given by
$\{|e_1,g_2,N-1\rangle,|g_1,e_2,N-1\rangle,|g_1,g_2,N\rangle\}$ while
for $N\ge 2$ a basis is given by $\{|e_1,e_2,N-2\rangle,
|e_1,g_2,N-1\rangle, |g_1,e_2,N-1\rangle, |g_1,g_2,N\rangle\}$.  The
third label indicates the number of photons. Although  the number of
excitations is conserved in both models (DJC and two-atom DD), the
main differences in the  time evolution of the system arise from the
collective atomic dynamics governing the quantum dynamics in  the DD
model, which is of course not present in the DJC situation.

\section{Entanglement and linear entropy}

Let us first consider the dynamics in the subspace with $N=1$
excitations. For the sake of simplicity, we consider the initial state
to be $|\Psi(0)\rangle=|e_1,g_2,0\rangle$, such that the unitary time
evolution of the whole system's state, under both DJC and DD
interaction, is given by
\begin{eqnarray}  |\Psi(t)\rangle=a_1(t)|e_1,g_2, 0\rangle + a_2(t)|g_1,e_2,0\rangle
+a_3(t)|g_1,g_2, 1\rangle
\label{eq:state}
\end{eqnarray}  This allows us to study the role of the cavity as as ``quantum data bus' to
 generate/transfer entanglement, as well as considering the cavity as
 a third qubit. Hence the atom-atom and single-atom-field
 entanglements  can be quantified using the concurrence
 \cite{wooters}.  We have found simple expressions  for these
 quantities:
\begin{eqnarray} C(\rho_{a,a})(t)&=&2|a^*_1(t)a_2(t)|\nonumber \\
C(\rho_{a1,f})(t)&=&2|a^*_1(t)a_3(t)|\\
C(\rho_{a2,f})(t)&=&2|a^*_2(t)a_3(t)|\nonumber
\end{eqnarray} The linear entropy of a particular sub-system ($A$) can be calculated using  the
relation $M_A=1-Tr{\rho_A^2}$, where $\rho_A$  is the partial trace of
$|\Psi(t)\rangle\langle\Psi(t)|$ over the $BC$  sub-system.  The
linear entropies for the field ($M_{f}$),  the atom$_1$ ($M_{a1}$),
and the atom$_2$ ($M_{a2}$) are given by
\begin{eqnarray} M_f&=&1-[(|a_1(t)|^2+|a_2(t)|^2)^2+|a_3(t)|^4]\nonumber \\
M_{a1}&=&1-[(|a_2(t)|^2+|a_3(t)|^2)^2+|a_1(t)|^4]\\
M_{a2}&=&1-[(|a_1(t)|^2+|a_3(t)|^2)^2+|a_2(t)|^4]\nonumber
\end{eqnarray} Under this unitary evolution, the  mixedness for the state of each element in the
system only arises from  entanglement with the other components: hence
we make the conjecture that  {\em the individual  linear entropy
should be a good measure of how entangled  each element is} with the
rest of the system.  Therefore an appropriate addition (subtraction)
of individual entropies  should be capable of quantifying the degree
of entanglement within each of  the sub-systems,
i.e. atom$_1$-atom$_2$, atom$_1$-field, and  atom$_2$-field. We
therefore define the {\em intrinsic} entanglement  within the $AB$
subsystem as
\begin{eqnarray} E_{A,B}=M_{A}+M_{B}-M_{A,B}
\label{eq:indicator}
\end{eqnarray} where $M_{A,B}$ indicates {\em how entangled (mixed)} the $AB$ sub-system is with 
the remaining constituents of the system. We compare the two-atom
concurrence  $C(\rho_{a,a})$ and the atom$_i$-field concurrences
$C(\rho_{ai,f})$  with the intrinsic entanglements $E_{a,a}$ and
$E_{ai,f}$. We find that these two measures for entanglement have
identical functional forms, namely:
\begin{eqnarray} E_{a,a}&=& M_{a1}+ M_{a2}-M_f  =  [C(\rho_{a,a})]^2\nonumber \\ E_{a1,f}&=&
M_{a1}+M_f-M_{a2}  =  [C(\rho_{a1,f})]^2\label{eq:subent}\\
E_{a2,f}&=& M_{a2}+M_f-M_{a1}  =  [C(\rho_{a2,f})]^2 \nonumber
\end{eqnarray} Notice that $M_{A,B}=M_C$ since the whole system is in a pure state. Some
important aspects that follow from these relations are:
\begin{itemize}
\item The definition in Eq.(\ref{eq:indicator}) and the relations
given in Eq.(\ref{eq:subent}) agree with general inequalities that
quantum entropies should satisfy \cite{nielsen}: In particular, the
subadditivity inequality ($M_{A,B} \leq M_A+M_B$) if the system is
considered as a mixed two-qubit state, and the strong subadditivity
inequality ($M_{\{A,B,C\}} +M_B\leq M_{A,B}+M_{B,C}$) if the system is
assumed  as a composite system of three elements $A_1$, $A_2$, and
field.
\item  The definition in Eq.(\ref{eq:indicator}) satisfies all the
relevant criteria for an entanglement  measure \cite{bruss02}: (a) it
is semipositive, i.e. $E_{A,B} \ge 0$ where the equality sign holds
for a  separable state, (b) $E_{A,B}(t)$ is a continuous function of
time, as shown in the next sections, and  (c) $E_{A,B}$ is invariant
and nonincreasing under local unitary operations in a similar way to
the quantum entropies defining it.
\end{itemize}

\vskip0.1in

In more general situations, such a simple relationship between  $E$
and the concurrence cannot be expected to hold. To illustrate this, we
will consider dynamical evolutions within the subspaces of higher
excitations. In the subspace with $N=n+1$ excitations, the system's
state at time $t$ is given by
\begin{eqnarray}  |\Psi(t)\rangle&=&b_0(t)|e_1,e_2,n-1\rangle + b_1(t)|e_1,g_2,
n\rangle\\\nonumber & & + b_2(t)|g_1,e_2,n\rangle + b_3(t)|g_1,g_2,
n+1\rangle \ \ 
\end{eqnarray} The cavity can no longer be considered as a qubit in this situation, hence it
makes sense to focus instead  on the relationship between the
atom-atom concurrence $C(\rho_{a,a;N>1})$ and the intrinsic
entanglement $E_{a,a;N>1}(t)$. In this case, we find that
$E_{a,a;N>1}(t)\geq C^2(\rho_{a,a;N>1})(t)$.  Explicitly we obtain
$E_{a,a;N>1}(t)=4|b_1(t) b_2(t)|^2 + 2|b_0(t)b_3(t)|^2$, while
$C(\rho_{a,a;N>1})(t)=2(|b_1(t) b_2(t)|-|b_0(t)b_3(t)|)$ if
$2|b_1(t) b_2(t)|>|b_3(t) b_0(t)|$ but is zero otherwise.  Although
we do not yet have a definitive statement of the conditions under
which a simple relationship as in Eq. (\ref{eq:subent}) will hold, we
believe that it is related to the complexity of the multi-partite
subspace within which the system's Hamiltonian can couple states.

In what follows we will focus in the initial state
$|\Psi(0)\rangle=|e_1,g_2,0\rangle$  and will discuss results for the
two interaction models, i.e. DJC and two-atom DD interaction.

\subsection{Transfer of entanglement in the DJC situation} For this sequential interaction, the
dynamics can be divided into two steps.  Before the first atom leaves
the cavity, i.e. $t\leq \tau_1$, the unitary  evolution is given by
\begin{eqnarray}  a_1(t)=\cos[\theta_1(t)]\nonumber \\ a_2(t)=0\\
a_3(t)=-i\sin[\theta_1(t)]\nonumber
\end{eqnarray} Once $A_1$ has left the cavity, i.e.  $t>\tau_1$, it is easy to show that
\begin{eqnarray}  a_1(t)=\cos[\theta_1(\tau_1)]\nonumber \\
a_2(t)=-\sin[\theta_1(\tau_1)]\sin[\theta_2(t)]\\
a_3(t)=-i\sin[\theta_1(\tau_1)]\cos[\theta_2(t]\nonumber
\end{eqnarray} where $\theta_i(t)=\int_{t_i}^{t}\gamma_i(t')dt'$ for $t_i\leq t \leq
\tau_i$. Using this sequential interaction scheme, it is possible to
produce  a maximally entangled two-atom  state as well as a hybrid
entangled atom-atom-cavity $W$-state.    By adjusting the interaction
times such that  $\theta_1(\tau_1^s)=\pi/4$ and
$\theta_2(\tau_2^s)=\pi/2$, a  maximally entangled atomic state is
obtained, which is separable from the field state, i.e.
$|\Psi(\tau_s)\rangle=(1/\sqrt 2)\{|e_1,g_2\rangle
-|g_1,e_2\rangle\}\otimes |0\rangle$. The $W$-state is  obtained when
$\theta_1(\tau_1^w)=\arccos{(1/\sqrt{3})}$ and
$\theta_2(\tau_2^w)=\pi/4$, i.e. $|\Psi(\tau_w)\rangle=(1/\sqrt
3)\{|e_1,g_2,0\rangle  -|g_1,e_2,0\rangle + i|g_1,g_2,1\rangle\}$.

\begin{figure}
\begin{center}
\resizebox{8cm}{!}{\includegraphics*{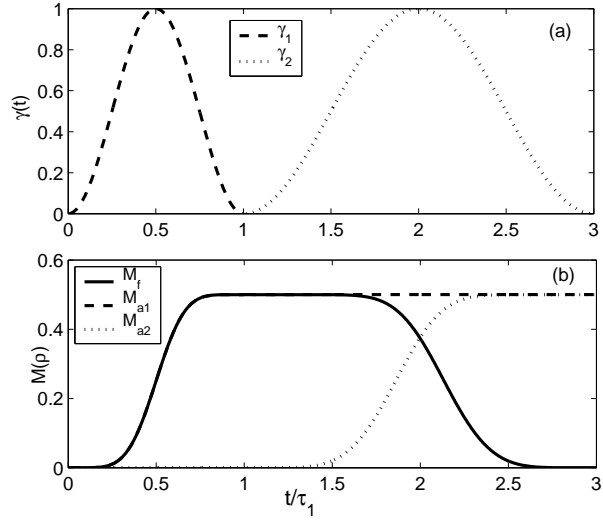}}
\caption{One-by-one DJC interaction during formation of a singlet
state.  (a)Coupling strengths and (b)individual linear entropies as
functions of time. Time in units of the interaction time corresponding
to $A_1$, $\tau_1$.}
\label{fig:jcsinent}
\end{center}
\end{figure}

\begin{figure}
\begin{center}
\resizebox{8cm}{!}{\includegraphics*{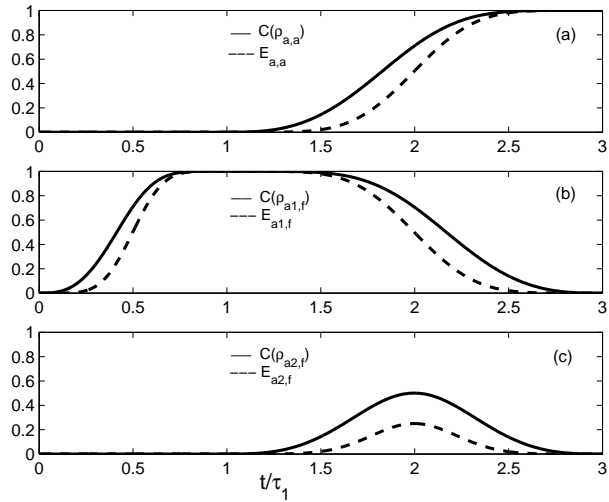}}
\caption{Dynamics of two-qubit entropies and concurrences during
generation of a  singlet state in the DJC model.(a) Atom-atom
intrinsic entanglement $E_{a,a}$ and concurrence $C(\rho_{a,a})$, (b)
$A_1-$field intrinsic entanglement $E_{a1,f}$ and concurrence
$C(\rho_{a1,f})$, and (c)$A_2-$field intrinsic entanglement $E_{a2,f}$
and concurrence  $C(\rho_{a2,f})$, as functions of time. Time is
measured in units of $\tau_1$.}
\label{fig:jcsincw}
\end{center}
\end{figure}

Figures \ref{fig:jcsinent} and \ref{fig:jcsincw} show, respectively,
the dynamical evolution of the individual entropies and the intrinsic
entanglements $E$ (as defined in Eq.(\ref{eq:subent})) during the
generation of the singlet state. As expected with this DJC
interaction, the combined state is pure while atom$_1$ is interacting
with the cavity. Hence the atom and field entropies increase
identically, i.e. $M_{a1}(t)=M_{f}(t)$ for $t \leq \tau_1$. The
maximum atom $A_1$-field entanglement is obtained when the individual
entropies reach their maximum value of 1/2 (as expected for a
maximally mixed state).

After the first atom leaves the cavity, its entropy $M_{a1}(\tau_1)$
remains constant ($A_1$ does not interact with the field anymore)
while $M_{a2}$ increases and $M_f$ decreases. An interesting interplay
between $M_{a2}$ and $M_f$ can be observed. In
Fig. \ref{fig:jcsinent}(b) it can be seen that there is a particular
time $t'$ when $M_{a2}(t')=M_f(t')$ and hence a {\em temporal}
symmetry arises: $M_{a2}(t'+\delta t)=M_f(t'-\delta t)$
with $\delta t \leq \tau_2$.  This is basically a manifestation of the
memory effects in the system by which atom $A_1-$field
entanglement is {\em converted into} atom-atom entanglement. This conversion can be
better appreciated by considering the dynamics of the intrinsic
entanglements (Fig.  \ref{fig:jcsincw}). A comparison of
Fig.\ref{fig:jcsincw}(a) and Fig.  \ref{fig:jcsincw}(b) allows us to
conclude that $E_{a,a}(t'+\delta)=E_{a1,f}(t'-\delta)$, with this
being achieved via $A_2$-field entanglement (see
Fig. \ref{fig:jcsincw}(c)). It is worth nothing that $t'$ is the
instant in time when $E_{a2,f}$ takes its maximum value (which is less
than one).

For the generation of the singlet, $M_f$ goes to zero while $M_{a2}$
approaches the constant value $M_{a1}(\tau_1^s)=1/2$. The fact that
$M_f(\tau_s)=0$ indicates that the field is in a pure state (i.e. it
is not entangled with either of the atoms). Hence atom-field
concurrences vanish as well as atom-field intrinsic entanglements
$E$. This effect can be seen in Fig. \ref{fig:jcsincw}. Since both
$M_{a1}(\tau_s)$ and $M_{a2}(\tau_s)$ approach the same finite maximum
value of $1/2$, the atom-atom concurrence and atom-atom entanglement
$E_{a,a}$ become equal one (as expected for a maximally entangled
two-qubit state).

\begin{figure}
\begin{center}
\resizebox{8cm}{!}{\includegraphics*{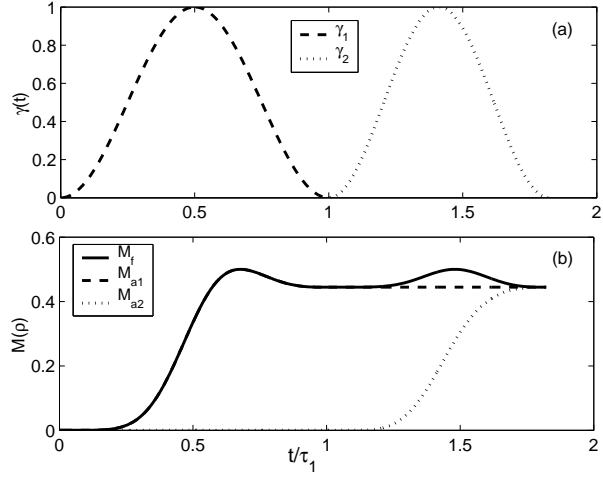}}
\caption{One-by-one DJC interaction during formation of a hybrid
atom-atom-cavity $W$-state. (a)Coupling strengths and (b)individual
linear entropies as functions of time. Time is measured in units of
$\tau_1$.}
\label{fig:jcwent}
\end{center}
\end{figure}

\begin{figure}
\begin{center}
\resizebox{8cm}{!}{\includegraphics*{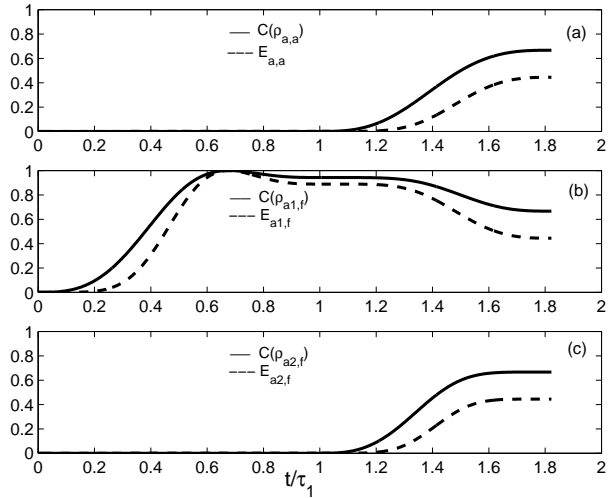}}
\caption{Dynamics of two-qubit entropies and concurrences while
forming  a hybrid atom-atom-cavity $W-$state in the DJC model.(a)
Atom-atom intrinsic entanglement $E_{a,a}$ and concurrence
$C(\rho_{a,a})$, (b) $A_1-$field intrinsic entanglement $E_{a1,f}$ and
concurrence  $C(\rho_{a1,f})$, and (c)$A_2-$field intrinsic
entanglement $E_{a2,f}$ and concurrence  $C(\rho_{a2,f})$, as
functions of time. Time is measured in units of $\tau_1$.}
\label{fig:jcwcw}
\end{center}
\end{figure}

When forming the W-state, the field and atom $A_1$ become maximally
entangled before the latter leaves the cavity. Hence
$M_{f}(\tau_1^w)<1/2$ as shown in Fig. \ref{fig:jcwent}(a).  After
$A_1$ leaves, $M_{a2}$ increases and $M_f$ varies non-monotonically
until they both equal $M_{a1}(\tau_1^w)<1/2$ (see
Fig.\ref{fig:jcwent}(b)), indicating that at $t=\tau_w$ the field and
atoms are identically mixed (entangled) with each other. All subsystem
concurrences approach the same value $2/3$ (as expected for a
$W$-state). A similar behaviour is observed for the subsystem
entanglements: at $t=\tau_w$ all of them equal $4/9$ (see
Fig. \ref{fig:jcwcw}).

\subsection{Sharing of entanglement in the two-atom DD interaction} 
\begin{figure}
\begin{center}
\resizebox{8cm}{!}{\includegraphics*{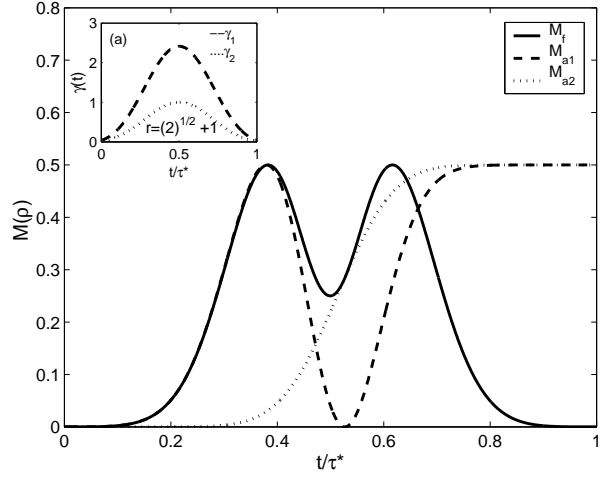}}
\caption{Dynamics of individual linear entropies during the  formation
of a triplet state $|\Phi^+,0\rangle$. (a)Coupling strengths as
functions of time.}
\label{fig:ddtrent}
\end{center}
\end{figure}

\begin{figure}
\begin{center}
\resizebox{8cm}{!}{\includegraphics*{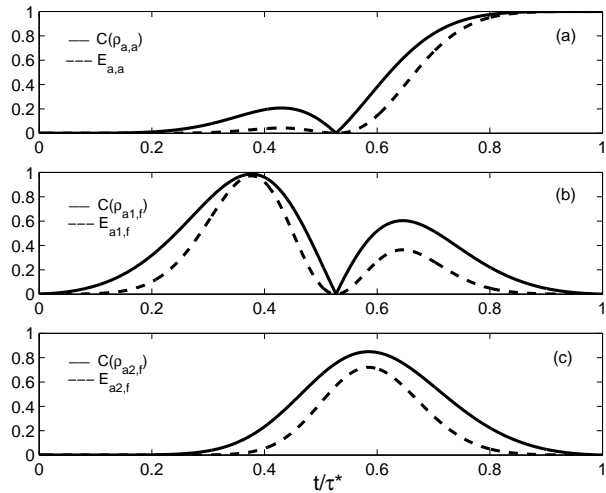}}
\caption{Dynamics of two-qubit intrinsec entanglement $E$ and concurrences 
while
forming  a triplet state $|\Phi^+,0\rangle$ in the two-atom DD
model.(a) Atom-atom entropy $E_{a,a}$ and concurrence $C(\rho_{a,a})$,
(b) $A_1-$field entropy $E_{a1,f}$ and concurrence  $C(\rho_{a1,f})$,
and (c)$A_2-$field entropy $E_{a2,f}$ and concurrence
$C(\rho_{a2,f})$, as a function of time.}
\label{fig:ddtrcw}
\end{center}
\end{figure}

We have previously
demonstrated that the use of different time-dependent and  asymmetric
on-resonance atom-field couplings, allows the generation of  two-qubit
states with any degree of entanglement and a prescribed symmetry
\cite{ojqpr03}.  This is achieved by exploiting a trapping vacuum
state condition, which does  not arise for identical couplings
\cite{ojqpr03}.  Here we focus on the situation in which the couplings
satisfy the relation  $\gamma_1(t)=r\gamma_2(t)$. With this condition,
we have found that the  unitary evolution for the two-atom DD model
with initial state  $|\Psi(0)\rangle=|e_1,g_2,0\rangle$ is given by
\cite{ojqpr03}:
\begin{eqnarray}  a_1(t)&=&1+ra_2(t)\nonumber \\  a_2(t)&=&-2\alpha{\rm
sin}^2[\theta(t)/2]\label{eq:coef}\\  a_3(t)&=&-i\sqrt{r \alpha} {\rm
sin}[\theta(t)]\nonumber
\end{eqnarray}  where $\theta(t)=\int_0^t \omega(t')dt'$ is the effective vacuum  Rabi angle. The
time-dependent frequency of the collective atomic mode  coupled  to
the cavity field is given by $\omega^2(t)=\gamma_1(t)^2
+\gamma_2(t)^2$,  while
$\alpha=\gamma_1(t)\gamma_2(t)/\omega^2(t)=r/(1+r^2)$ denotes the
relative  geometric mean of the couplings. From Eqs. (\ref{eq:state})
and (\ref{eq:coef}) it  is clear that if $r\neq1$, there is a
particular time $\tau^*$ when  $\theta(\tau^*)=\pi$ such that  the
atom-field  state is separable, but the  two-atom state remains
entangled:
\begin{eqnarray} |\Psi(\tau^*)\rangle=[a_1(\tau^*)|e_1,g_2\rangle +
a_2(\tau^*)|g_1,e_2\rangle]\otimes|0\rangle \ \ .
\label{eq:entangled}
\end{eqnarray} When the condition $a_1(\tau^*)=\pm a_2(\tau^*)=\pm 1/\sqrt 2$ holds,  leading to
$r_{\pm}=\sqrt 2 \pm 1$, the two-atom state becomes maximally
entangled, i.e.  $|\Phi^{\pm}\rangle=[|e_1,g_2\rangle \pm
|g_1,e_2\rangle]/\sqrt{2}$.  When the couplings are identical the
separability of atomic and field states  implies separability in the
atom-atom states as well. This is due to the  fact that in this
simultaneous interaction scheme,  identical atom-field  couplings lead
to a  symmetric entropy dynamics for both atoms. By contrast,
asymmetric couplings lead to a more complex dynamics. To demonstrate
this,  we now discuss the dynamics of the individual entropies (see
Fig.\ref{fig:ddtrent}) and intrinsec entanglements (see 
Fig.\ref{fig:ddtrcw})
while  forming the maximally entangled triplet state
$|\Phi^+\rangle=[|e_1,g_2\rangle + |g_1,e_2\rangle]/\sqrt{2}$
$(r=\sqrt 2 +1)$. At the beginning of the  interaction, the initially
excited atom $A_1$ is the most strongly coupled  to the cavity field
$(r>1)$. Hence its entropy $M_{a_1}$ increases faster  than
$M_{a_2}$. This manifests itself in the fact that both the $A_1-$field
concurrence $C(\rho_{a_1,f})$ and entanglement  $E_{a_1,f}$
(Fig. \ref{fig:ddtrcw}(b)) grow faster than  $C(\rho_{a2,f})$ and
$E_{a2,f}$ (Fig. \ref{fig:ddtrcw}(c)), respectively.  They follow
non-monotonic  evolutions in such a way that there is a time when atom
$A_1$ is not  entangled either with the field or with atom
$A_2$. Hence its linear entropy $M_{a_1}=0$
(Fig. \ref{fig:ddtrent}(b)). At  the end of the interaction period
$(t=\tau^*)$, both $C(\rho_{a1,f})$ and $C(\rho_{a2,f})$ vanish while
the  atom-atom concurrence and intrinsic entanglement go to the
maximum value of  unity (see Fig. \ref{fig:ddtrcw}(a)).
 
We finish by discussing how the present study might be used to develop
correlation measures for multi-partite systems. Our results have shown
that  the additivity of the linear entropy seems to have  a clear
physical meaning in the particular cases considered here, where the
entire system can be considered as a multi-qubit system
(i.e. three-qubit in the present case).  We have  shown that by
properly adding linear entropies we can define an alternative  measure
for two-qubit entanglement which has the same  functional form as the
concurrence, and satisfies the relevant criteria for  such a measure.
In a similar way, one might be able to extend this analysis to
many-particle systems by studying the dynamics of quantum correlations
in $N$-particle mixed states under unitary evolutions, i.e. in a
$N+1-$particle entangled state. In particular it has recently been
shown for $N=3$, based on the strong subadditivity inequality for von
Neunman entropies, that one can define a possible measure for
four-particle entanglement\cite{biswas}.  By studying such
inequalities for linear entropies, one might gain deeper insights into
the open problem concerning entanglement measures for many-particle
states.  An extension of the present study to the case of $N\ge 3$
qubits interacting with a cavity mode, is underway and will be
reported elsewhere. An interesting first step would be to investigate
whether a generalization of our definition of intrinsic entanglement
$E$ holds for the case of  an $N$-qubit system $\{a_1,a_2,...,a_N\}$
in a pure quantum state $|\Psi_N\rangle$. Specifically, one could
investigate  whether  $E$  for the $\{a_1, a_2,...,a_m\}$ sub-system
is given by
$E_{\{a_1,...,a_m\}}=\sum_{i=1,m}M_{a_i}-M_{\{a_1,...,a_m\}}$, or
whether this might instead become an inequality\cite{vedral03}. 

\section{Summary} We have found quantitative relations between entanglement
and linear entropy  for a two-qubit-cavity system governed by unitary
evolutions. By considering  the behaviour of individual sub-system's
entropies, we have been able to  propose a measure of bipartite
entanglement. Furthermore, we have discussed  how these relations
might be extended to multi-partite systems. In parallel  with future
extensions to many-particle systems, it will be interesting to
consider the full effects of open systems with regards to
decoherence. In  particular, it will be of fundamental interest to
understand the role of the  environment with regards to mixedness,
entanglement and entropy. The  atom-cavity quantum optics systems of
the type studied here, should provide  ideal theoretical and
experimental systems for pursuing such studies.

\ack We acknowledge funding from the Clarendon Fund and ORS (AOC),
DT-LINK (NFJ) and COLCIENCIAS project 1204-05-13614 (LQ).\\


\end{document}